\documentclass{ws-ijmpd}
\usepackage{savesym}
\savesymbol{captionbox}
\usepackage{subcaption}
\restoresymbol{TMP}{captionbox}
\usepackage[super,compress]{cite}
\usepackage{url}

\begin{document}

\markboth{Makukov and Mychelkin}
{Rotation in vacuum and scalar background}

%
\catchline{}{}{}{}{}
%

\title{ROTATION IN VACUUM AND SCALAR BACKGROUND:\\ ARE THERE 
ALTERNATIVES TO NEWMAN-JANIS ALGORITHM?}

\author{MAXIM MAKUKOV \and EDUARD MYCHELKIN}

\address{Fesenkov Astrophysical Institute, 050020, Almaty, Republic of
Kazakhstan\\
makukov@fai.kz, mychelkin@fai.kz
}

\maketitle

\begin{abstract}

The Newman-Janis algorithm is the standard approach to rotation in
general relativity which, in vacuum, builds the Kerr metric from the
Schwarzschild spacetime. Recently, we have shown that the same
algorithm applied to the Papapetrou antiscalar spacetime produces a
rotational metric devoid of horizons and ergospheres. Though exact in
the scalar sector, this metric, however, satisfies the Einstein
equations only asymptotically. We argue that this discrepancy between
geometric and matter parts (essential only inside gravitational radius
scale) is caused by the violation of the Hawking-Ellis energy
conditions for the scalar energy-momentum tensor. The axial potential
functions entering the metrics appear to be of the same form both in
vacuum and scalar background, and they also coincide with the
linearized Yang-Mills field, which might hint at their common
non-gravitational origin. As an alternative to the Kerr-type
spacetimes produced by Newman-Janis algorithm we suggest the exact
solution obtained by local rotational coordinate transformation from
the Schwarzschild spacetime. Then, comparison with the Kerr-type
metrics shows that the Lense-Thirring phenomenon might
be treated as a coordinate effect, similar to the Coriolis force. 

\end{abstract}


\ccode{PACS numbers:}


\section{Introduction}

The Kerr solution \cite{kerr_gravitational_1963} of the vacuum
Einstein equations (EE) represents a mainstream approach to the
rotation problems in general relativity (GR). Still, there are certain
aspects about this metric which remain problematic. In spite of the
variety of efforts undertaken from the outset
\cite{krasinski1978ellipsoidal}, the gravitating source for the Kerr
metric is still unknown \cite{Padmanabhan2010}. Besides, absence of
the interior solution, existence of closed time-like curves and
the very stability of the Kerr solution are among important unresolved
issues in GR \cite{andersson_nonlinear_2022}. In view of this, it is
worthwhile to look for alternative approaches, and one of those
is to consider the gravitating scalar field as realistic background
instead of vacuum.

The Newman-Janis (NJ) algorithm \cite{Newman1965} was first applied in
Refs.~ \refcite{Krori1981} and \refcite{Agnese1985} to the general static
Janis-Newman-Winicour (JNW) family of solutions \cite{Janis1968} of
the EE with massless scalar field parameterized by the factor
$\gamma$ related to scalar charge and taking the values  within
the range $[0,1]$. However, as had been recently established by the
method of successive approximations \cite{bogush_galtsov20}, the
result of such application of the NJ algorithm does not satisfy the
Einstein-scalar equations. Meanwhile, the JNW-type spacetimes, starting with the pioneering
attempt by Fisher to find the closed solution in curvature coordinates
\cite{fisher}, are, by themselves, unstable with respect to collapse,
and thus do not seem to be physically relevant
\cite{abe1988stability,christodoulou1999instability,liu2018robust,faraoni2021spherical}.

Recently, as an alternative to the JNW spacetimes, we have
investigated \cite{mm18} the antiscalar (i.e., having the opposite
sign of the scalar field energy-momentum tensor) exponential
Papapetrou solution \cite{papa} and revealed that not only it produces
practically the same observational effects as in vacuum case but is
also characterized by thermodynamic stability as well as the
Klein-Gordon equation stability. Moreover, it leads to the
value of scalar charge (the source of scalar field) being equal to the
central mass \cite{makukov_triple_2020}. Thus, the application of the
NJ algorithm to the Papapetrou metric (resulting in what we call the
Papapetrou-Newman-Janis metric, or PNJ for short) might be of
interest. Since the exponential metric is functionally related to the
JNW solution via the limit $\gamma \to \infty$
\cite{mm18,makukov_triple_2020}, which falls out of the $\gamma$-range
mentioned above, the problem of compatibility of the PNJ metric with
the EE requires a separate study, which is one of the aims in this
paper. Given that the source for the Kerr-type metrics is not
established, we will also consider an alternative approach based on
exact rotational solutions derived with direct coordinate
transformation from the Schwarzschild metric.

\section{Setup}%
\label{sec:setup}

The application of the complex-shifting NJ algorithm to the Schwarzschild metric 
\begin{equation}
ds^2 =  (1-2M/R)dt^2 - (1-2M/R)^{-1} {dR}^2 -  R^2 d\Omega^2,
\label{Schwarz}
\end{equation}
leads to the Kerr solution (for nomenclature reasons, we also call it
the Schwarzschild-Newman-Janis or SNJ metric):
\begin{eqnarray}
ds^2 &=& (1-2MR/\rho^2) \left(dt- a\sin^2\theta d\phi \right)^2 \nonumber\\
&-& \rho^2\left(\frac{dR^2}{\Delta}+d\theta^2+\sin^2\theta d\phi^2\right)\nonumber\\
&+& 2a\sin^2 \theta(dt- a \sin^2\theta d\phi)d\phi,
\label{kerr}
\end{eqnarray}
where $\Delta = R^2  + a^2  - 2M R$. Note that the $g_{00}$ component contains the scalar function 
\begin{equation}
\varphi(R,\theta,a)=MR/\rho^2,  \quad   \rho^2(R) = R^2 +a^2 \cos^2\theta,
\label{phiKerr}
\end{equation}
which results from complexifying the radial coordinate (see below), with the meaning of the parameter $a$ left open within the NJ algorithm.

The exponential Papapetrou metric 
\begin{equation}
ds^2 =  e^{-2\varphi(r)} {dt}^2 - e^{2\varphi(r)} \left( {dr}^2 + r^2 d\Omega^2
\right),
\label{Papa}
\end{equation}
represents a static spherically symmetric solution of the general Einstein-scalar field equations
\begin{equation} 
    {G}_{\mu\nu} = \epsilon \varkappa {T}_{\mu\nu}^{\text{sc}}\left( \varphi \right) ,
\label{EinEq}
\end{equation}
in antiscalar regime, i.e. with $\epsilon = -1$. Here, $\varkappa=8 \pi G /c^4$ (we use $G=c=1$) and the minimal scalar energy-momentum tensor (EMT)
\begin{equation}
    {T}^{\text{sc}}_{\mu\nu}(\varphi) = \frac{1}{4 \pi} \left( \varphi_\mu
    \varphi_\nu - \frac{1}{2} {g}_{\mu\nu} \varphi^\alpha \varphi_\alpha
\right), \,\,\,\, \varphi_\mu \equiv \partial_\mu \varphi,
\label{EMT}
\end{equation}
The values $\epsilon = 0$ and  $\epsilon =
1$ correspond to vacuum (Schwarzschild) and
scalar (JNW) solutions, correspondingly.
Taking the concomitant Klein-Gordon equation into account, it follows exactly that $\varphi(r)= M/r$
for the Papapetrou metric \eqref{Papa}.

The result of application of the NJ algorithm to the Papapetrou
solution is the PNJ metric \cite{mm18}:
\begin{multline}
  ds^2 = e^{-2\varphi(r,\theta,a)} \left(dt- a\sin^2\theta d\phi \right)^2 
       - e^{2\varphi(r,\theta,a)} \rho^2\left(\frac{dr^2}{\underline{\Delta}}+d\theta^2+\sin^2\theta d\phi^2\right)  \\
+ 2a\sin^2 \theta(dt- a \sin^2\theta d\phi)d\phi.
\label{rotPap}
\end{multline}
This metric is similar in structure to the Kerr solution, but unlike
in \eqref{kerr} here $\Delta \rightarrow\underline{\Delta}=r^2 + a^2$,
i.e. in this metric there are no horizons and ergospheres. For $a = 0$
the expression \eqref{rotPap} reduces to the Papapetrou solution \eqref{Papa}. 

The Klein-Gordon equation for the metric \eqref{rotPap} leads to
stationary rotational potential (which in this case is related to a
physical scalar field) of the same functional form as in
\eqref{phiKerr}:
\begin{equation}
g^{\mu\nu}\varphi_{;\mu\nu}=0 \,\, \Rightarrow \,\,  \varphi=\varphi(r,\theta,a) =  \frac{M r}{\rho^2} = \frac{M r}{r^2 +a^2 \cos^2\theta},
\label{potRot}
\end{equation}
which in the absence of rotation ($a=0$) reduces to the Newtonian
form. Note that the isotropic coordinate $r$ here should not be
confused with the curvature coordinate $R$.

We use the term \textit{physical} here in contrast to the Kerr,
Newman-Janis and Kerr-Schild approaches where the same potential
function enters into corresponding vacuum spacetime as a part of the
metric coefficients after some \textit{mathematical} manipulations. In our case
the zero four-divergence of physical scalar EMT (6) being equivalent
to the Klein-Gordon equation leads (in given coordinates) to exactly 
the same functional dependence for the physical background scalar field.

It is essential that final expressions \eqref{rotPap}, \eqref{potRot}
are obtained without using the Einstein equations and so, being
related directly to the scalar sector of the Lagrangian, might be
considered as self-consistent. The resulting (quite
cumbersome\footnote{We use Wolfram Mathematica to derive analytical
expressions, with sanity checks (verifying limiting cases, etc.)}) form of
the scalar EMT follows directly from substitution of \eqref{potRot}
and \eqref{rotPap} into \eqref{EMT}, with the only non-zero
off-diagonal components being $T_{r \theta}^{sc}$ and $T_{t
\phi}^{sc}$.

\section{Role of scalar background}
\label{sec:essence}

In our approach, the expression \eqref{EMT} used in \eqref{EinEq}
represents a fundamental field irremovable from the field equations
\cite{mm18,makukov_triple_2020}, and this imposes certain constraints
onto the theory as a whole, and the rotation problem in particular.

In general, the Einstein tensor and EMT, $G_{\mu\nu}$ and $\kappa
T_{\mu\nu}$, are distinct objects subjected to a strict constraint in
the form of the contracted Bianchi identity ${G^{\mu\nu}}_{;\nu}\equiv
0$, which represents a necessary but certainly insufficient condition
for the justification of GR as a closed theory of gravitational field. 

The reason for GR not being a closed theory is that the Ricci scalar
$R={R^\alpha}_\alpha$ used in the Hilbert-Einstein action as geometric
(`gravitational') part of the Lagrangian does not have the canonical
(Yang-Mills-like gauge-invariant) structure quadratic with respect to
first derivatives of a well-defined physical field. This circumstance
leads, in particular, to such obstacle as the absence of the EMT for
gravitational field itself (pseudo-tensors do not, strictly speaking,
solve the problem). Taking scalar background into account may
alleviate some of these problems.

At the same time, the EE prove to be an excellent tool for the
description of static configurations and dynamics of homogeneous
systems. In this respect note that transfer from the static metric to
stationary (rotational) regime is nontrivial already in vacuum.
Nontriviality follows, in particular, from the fact that it is not
known in advance if the applied NJ algorithm will conserve the
Ricci-flat character of the field equations. 

In scalar background the problem of such transfer (generated by the NJ or
some other algorithm) is much more complicated because it leads to
specific transformations for both sides of the EE, which are quite
different functionals with respect to metric. It is not obvious at all
that coincidence of those in static cases should imply their
compatibility in stationary regime as well. 

The applicability of some EMT within the EE might be tested with the
Hawking-Ellis (HE) approach \cite{Hawking1973} based on corresponding
eigenvalue problem, i.e.
\begin{equation}
\left( T_{\alpha\beta} - \lambda g_{\alpha\beta} \right) V^\beta = 0,
\end{equation}
with preliminary diagonalization of the EMT via local Lorentz
rotations implied, if needed. Considering the conformity with various
energy dominance conditions (for more details see also Ref.~\refcite{VisserHE17}), this approach leads to only four classification
types for EMTs compatible with the EE. The predominant type
I has one timelike and three spacelike eigenvectors. For contravariant
components it is expressible in diagonal form,
\begin{equation}
T^{\alpha\beta} \sim \left(
\begin{matrix}  
\lambda_1 & 0 & 0 & 0 \\
0 & \lambda_2 & 0 & 0 \\
0 & 0 & \lambda_3 & 0 \\
0 & 0 & 0 & \lambda_4 \\
\end{matrix}\right),
\label{HE}
\end{equation}
with all admissible eigenvalues $\lambda_\mu$ being real. We will not
consider the other three types being unstable or violating some of the energy conditions.

As a matter of fact, only for definite types of metrics, when the
symmetries found within the structure of the EMT are found in the
Einstein tensor as well, we have meaningful integrable systems such as
perfect fluids, static systems obeying time-like Killing field,
homogeneous expanding universe models ({\it a la} Friedmann), and, as
a degenerate case, the Ricci-flat vacuum.

In particular, as noted in Ref.~\refcite{galloway_topology_2021}, in
irrotational case, i.e. for vanishing vorticity $\omega=0$, the new
symmetry conditions for $u$-congruence, $u\wedge du=0$, together with
the Frobenius theorem guarantee integrability for a wide class of
systems (including static scalar background) within corresponding
$(1+3)$-splitting approach. 

The minimal NJ method, based on {\it linear} complex shift of
four-dimensional coordinates, is unique in a sense that it conserves
the Ricci-flat character of equations. Indeed, in static limit the
standard Kerr solution reduces to the vacuum Schwarzschild spacetime
in curvature coordinates which, due to the `reciprocal' condition
\begin{equation}
g_{00}g_{11}= -1, 
\label{toLin}
\end{equation}
forces the Einstein equations to be {\it linear}
\cite{Padmanabhan2010}. Exactly this linearity leads to translation
invariant character of the field equations and, as an exclusive
feature, to the Ricci flatness-preserving action of the NJ algorithm.

But in case of the scalar background such linearity is absent and,
besides, behavior and structure of the EE in static and stationary
regimes are cardinally different. It is remarkable that in the PNJ
case all analytical results can be obtained exactly, and, as will be
shown, the mismatch, or ``splitting'', between the Einstein tensor and the scalar EMT
proves to be essential only for small distances inside gravitational
radius scale.

\section{Stationary vs. static cases}%
\label{sec:genuine_reason}

In rotational systems the existence of additional axial Killing field,
$K=\partial_t + \omega \partial_\phi$ (see Ref.~\refcite{mm18}), leads to the
conservation of energy and angular momentum, but does not guarantee
the compliance with the Hawking-Ellis energy conditions. Indeed,
direct substitution of the PNJ metric \eqref{rotPap} with
\eqref{potRot} into general relation \eqref{EMT} leads to the EMT
non-diagonalisable by local Lorentz transformations. This violates the
Hawking-Ellis condition \eqref{HE}, and, as a result, certain
discrepancies arise between the left and right sides of the EE.

\begin{figure}
    \centering
    \begin{subfigure}{0.6\textwidth}
        \caption{Covariant components} \label{fig:splitting1}
        \includegraphics[width=\linewidth]{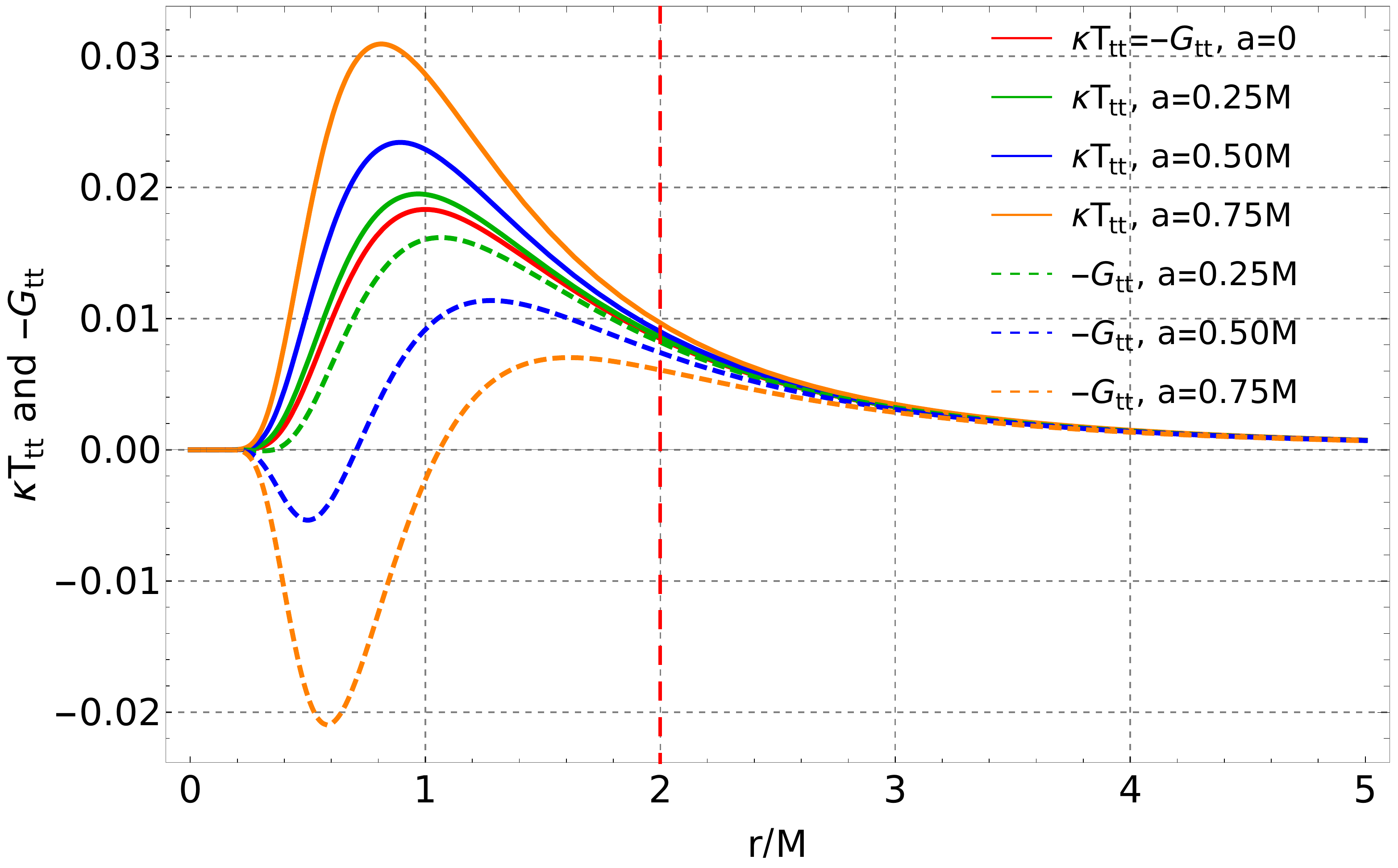} 
    \end{subfigure}
    \begin{subfigure}{0.6\textwidth}
        \caption{Mixed components} \label{fig:splitting2}
        \includegraphics[width=\linewidth]{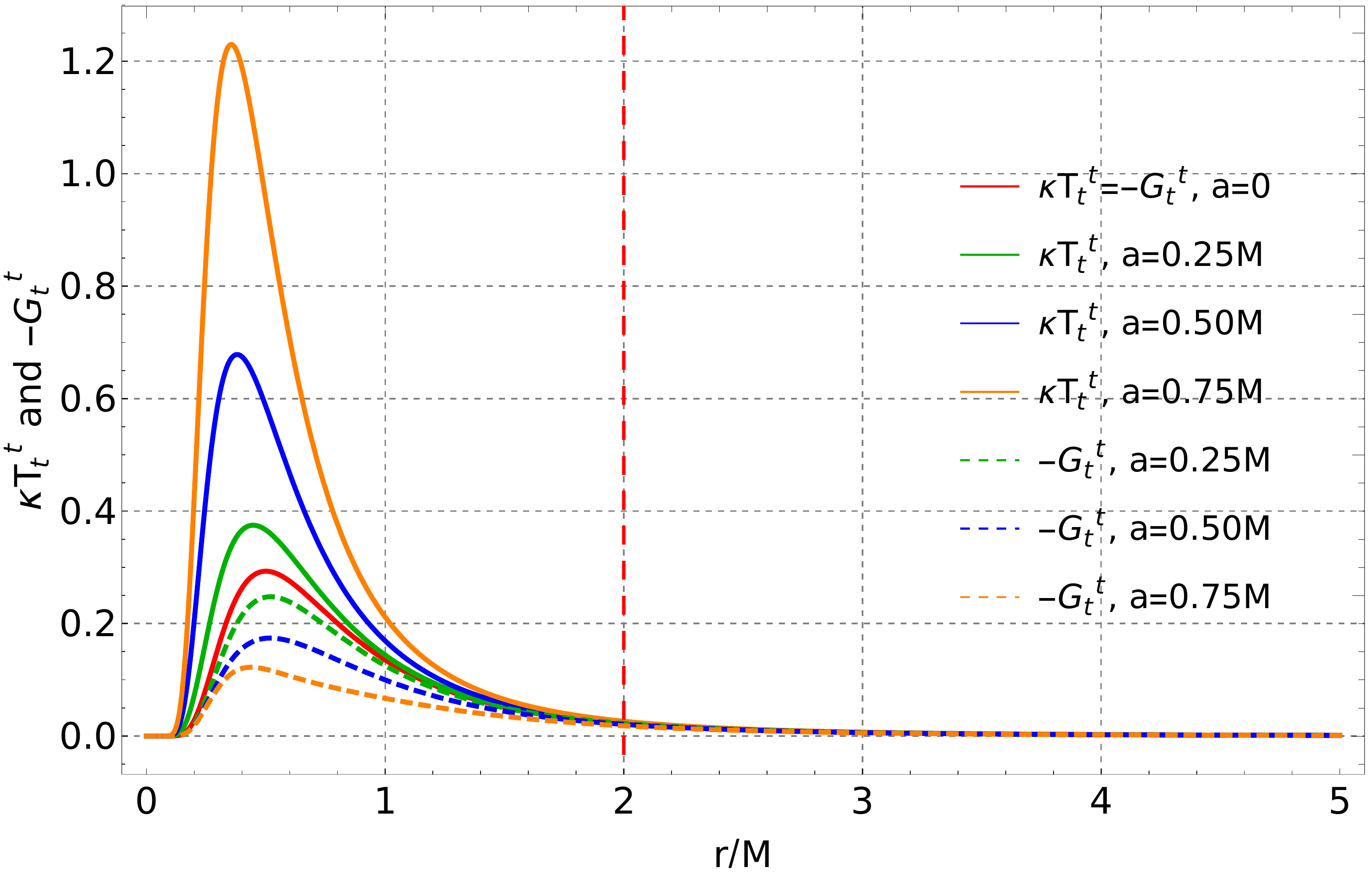} 
    \end{subfigure}
    \begin{subfigure}{0.6\textwidth}
        \caption{Contravariant components} \label{fig:splitting3}
        \includegraphics[width=\linewidth]{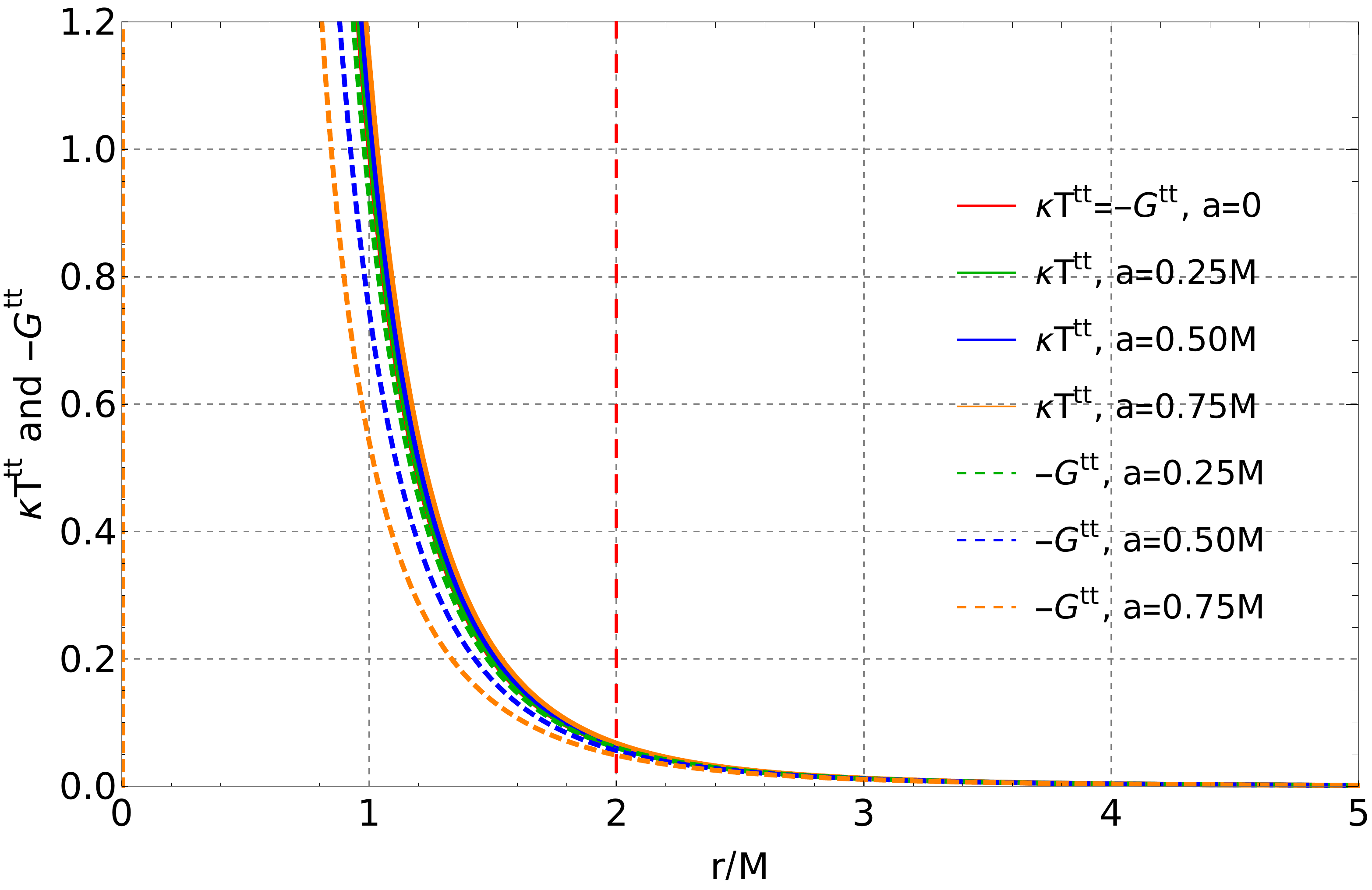} 
    \end{subfigure}
    \caption{Splitting between covariant (\subref{fig:splitting1}), mixed
        (\subref{fig:splitting2}), and contravariant
        (\subref{fig:splitting3}) $tt$-components of the EMT times
        $\varkappa$ (solid lines) and the Einstein tensor taken with
        negative sign (dashed lines) for various values of $a$ in the PNJ
        metric, for $\theta=\pi/2$. In all cases, the curves for the
        Einstein tensor go below the static red line ($a=0$), and at small
        distances the covariant components change sign. The splitting is
        essential only at scales $r < r_g=2M$ (red dashed line).}
\end{figure}

Specifically, taking the equatorial plane and designating the static
case $a=0$ with solid red lines in
Figs.~\ref{fig:splitting1}-\ref{fig:splitting3}, the following
regularities can be observed. While covariant component $T_{00}$
increases in its absolute value, $G_{00}$ decreases with the growth of
rotation parameter $a$ and, simultaneously, at small 
distances ($r~\ll~r_g$) swaps the sign, which is forbidden as per energy conditions.
As follows from Section~\ref{sec:essence}, origin of such anomalies in
behavior of $G_{\mu\nu}$ for stationary scalar background can be
attributed to non-canonical character of the geometrical Lagrangian
(cf. Ref.~\refcite{muench_brief_1998}).

At the same time from Fig.~\ref{fig:splitting1} one can see that for
$r>r_g$ (as well as for small $a$) the distinction between the left
and right sides in the (antiscalar) EE \eqref{EinEq} becomes
asymptotically negligible. Such asymptotic behavior might analytically
be evident, in particular, from relations \eqref{appGtt} and
\eqref{appTtt} in Appendix~\eqref{appendix:gtt}. Variation of $\theta$
does not incur essential influence on splitting effect.

Note, that only mixed components of the EMT have (for $\theta =
\pi/2$) diagonal form
\begin{equation}
{T_\mu}^\nu|_{\theta=\frac{\pi}{2}} = \frac{M^2 \left(a^2+r^2\right) e^{-\frac{2 M}{r}}}{8 \pi  r^6}   \times \text{diag}(1,-1,1,1).
\end{equation}
Then, as shown in Fig.~\ref{fig:splitting2}, for mixed time-time component
as well as for time-time component of the contravariant (non-diagonal)
EMT (see Fig.~\ref{fig:splitting3}),
\begin{equation}
T^{tt} |_{\theta=\frac{\pi}{2}} = M^2\frac{a^2 e^{-4M/r} \left(2 e^{2M/r}-1\right)+r^2}{8 \pi  r^6}
\end{equation}
with the corresponding (negative) Einstein tensor time-time components
$-{G_t}^t$ and $-G^{tt}$ (not written for brevity), the splitting
effect does not exhibit the peculiar behavior with the change-of-sign
anomalies in the Einstein tensor. It might also be seen from
Fig.~\ref{fig:splitting3} that contravariant EMT components (which
enter into the HE conditions) all differ only little from the case
$a=0$ which satisfies the HE conditions.

We have considered for brevity the case of $tt$-components only, but
the ``splitting'' of the EE occurs in a similar fashion for other
components as well.

\section{Lense-Thirring effect for vacuum and PNJ metrics}

Here we do not consider weak field approximation but deal with
exact relations in strong field regime. The general  frequency
$\Omega$ of a test gyro in an arbitrary stationary spacetime with a
timelike Killing vector $K$ can be expressed in terms of differential
forms \cite{Chakraborty2017a} as
\begin{equation}
	\tilde{\Omega} = \frac{1}{2 K^2}\ast \left(\tilde{K} \wedge d\tilde{K}  \right),
\label{oneforms}
\end{equation}
where $\tilde{\Omega}$ and $\tilde{K}$ are the one-forms of $\Omega$
and $K$, and $\ast$ represents Hodge dual. The $K$ vector might be
represented as a linear combination of time-translational and
azimuthal vectors, $K = \partial_t +\omega\partial_\phi,$ where
$\omega$ is the angular velocity for an observer moving along integral
curves of the $K$-field \cite{Chakraborty2017a}.

For special case $\omega=0$ the vector form of the coordinate-free
general spin precession rate \eqref{oneforms} after direct but
cumbersome transformations reduces to the exact expression describing the
Lense-Thirring effect (for details see, e.g., Ref.~\refcite{Chakraborty2017a}):
\begin{multline}
\vec{\Omega}|_{\omega=0} = \vec{\Omega}_{LT} = \sqrt{-g_{rr}}\Omega_{LT}^r \hat{r} + \sqrt{-g_{\theta\theta}} \Omega_{LT}^\theta \hat{\theta}=  \\
=  \frac{1}{2\sqrt {-g}}\left[\sqrt{-g_{rr}}\left(g_{t\phi,\theta}
-\frac{g_{t\phi}}{g_{tt}} g_{tt,\theta}\right)\hat{r} \right. 
 -\left. \sqrt{-g_{\theta\theta}}\left(g_{t\phi,r}-\frac{g_{t\phi}}{g_{tt}}
g_{tt,r}\right)\hat{\theta}\right].
\label{GenPrec0}
\end{multline}
In non-relativistic limit this leads to known expression for the Lense-Thirring precession
\cite{Chakraborty2017a}. The magnitude of this vector is
\begin{equation}
	\Omega_{LT}(r,\theta) = |\vec{\Omega}_{LT}| = \sqrt{  -g_{rr} (\Omega_{LT}^r)^2 - g_{\theta\theta} (\Omega_{LT}^\theta)^2 },
	\label{magnitudeLT}
\end{equation}
which is used in subsequent calculations. For the Kerr metric
(\ref{kerr}) this becomes \cite{mm18}:
\begin{equation}
	\Omega^{Kerr}_{LT}(R, \theta)  =  \frac{a M}{\rho^3 \left(\rho^2-2 M R\right)}
	  \sqrt{4 R^2 \Delta \cos^2\theta +\left(\rho^2-2 R^2\right)^2\sin^2\theta }\,\,.
	\label{LTkerr}
\end{equation}

Now, going to scalar background for the PNJ metric \eqref{rotPap}, one obtains
that the vector \eqref{GenPrec0}  has, in accord with 
\eqref{magnitudeLT}, the magnitude:
\begin{multline}
	\Omega^{PNJ}_{LT} = \frac{a e^{-\phi(r,\theta)}}{ \rho^{5}\sqrt{r^2 + a^2}}
	\left\{
	\cos^2\theta \left[\left(1-e^{-2\phi(r,\theta)}\right) \rho^4+2
a^2 M r \sin^2\theta \right]^2 \right. + \\ \left. M^2 (r^2 + a^2) \left(\rho^2-2 r^2\right)^2 \sin^2\theta \right\}^{\frac{1}{2}},
	\label{LT_RAS}
\end{multline}
with $\phi(r,\theta)=Mr/\rho^2$ and $\rho(r,\theta)$ defined
by \eqref{potRot}.

\begin{figure*}[t!]
    \centering
    \begin{subfigure}[t]{0.5\textwidth}
        \centering
        \caption{vacuum (Kerr), $\theta=0$}
        \includegraphics[width=2.6in]{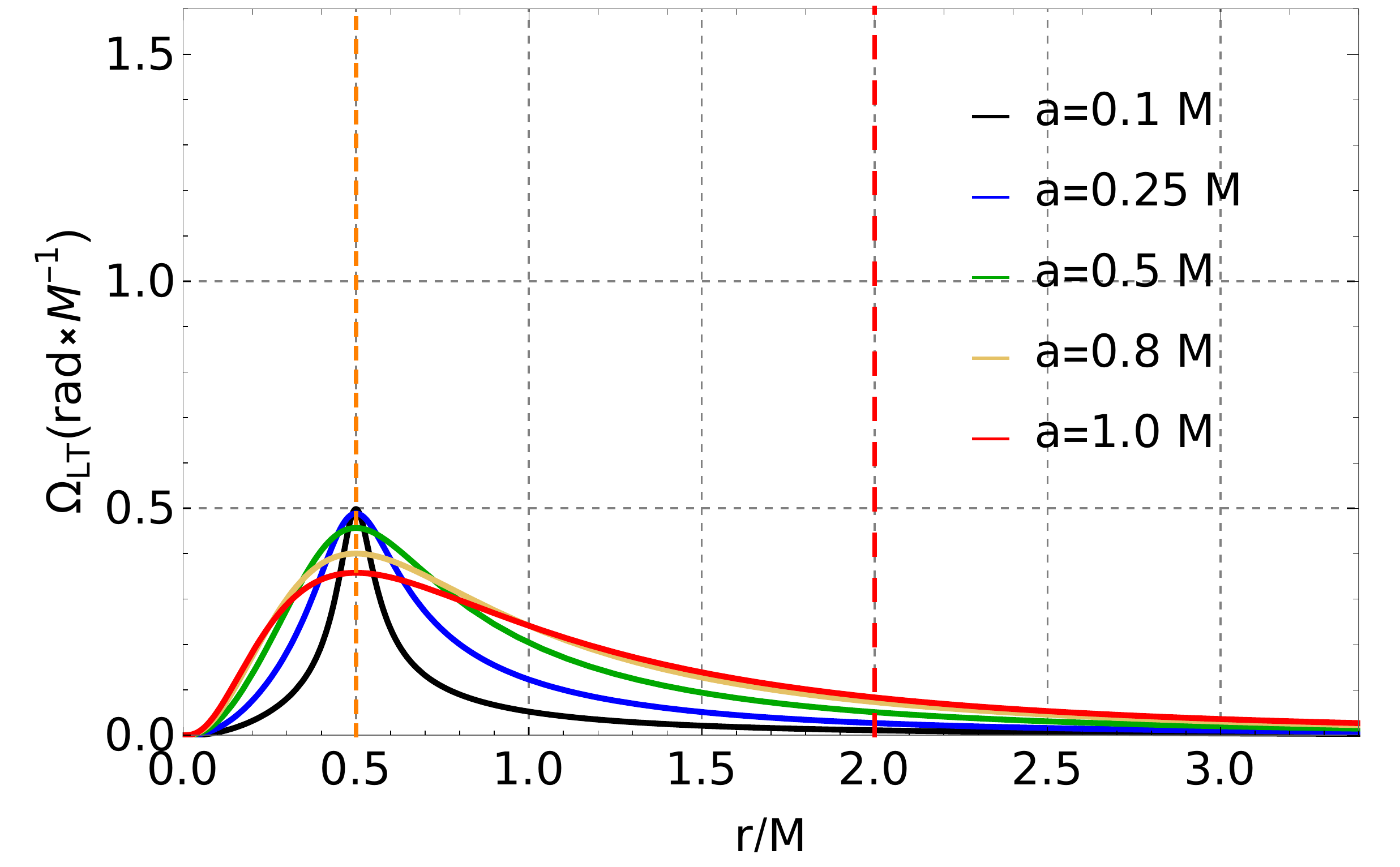}
    \end{subfigure}%
    ~ 
    \begin{subfigure}[t]{0.5\textwidth}
        \centering
        \caption{vacuum (Kerr), $\theta=\pi/2$}
        \includegraphics[width=2.6in]{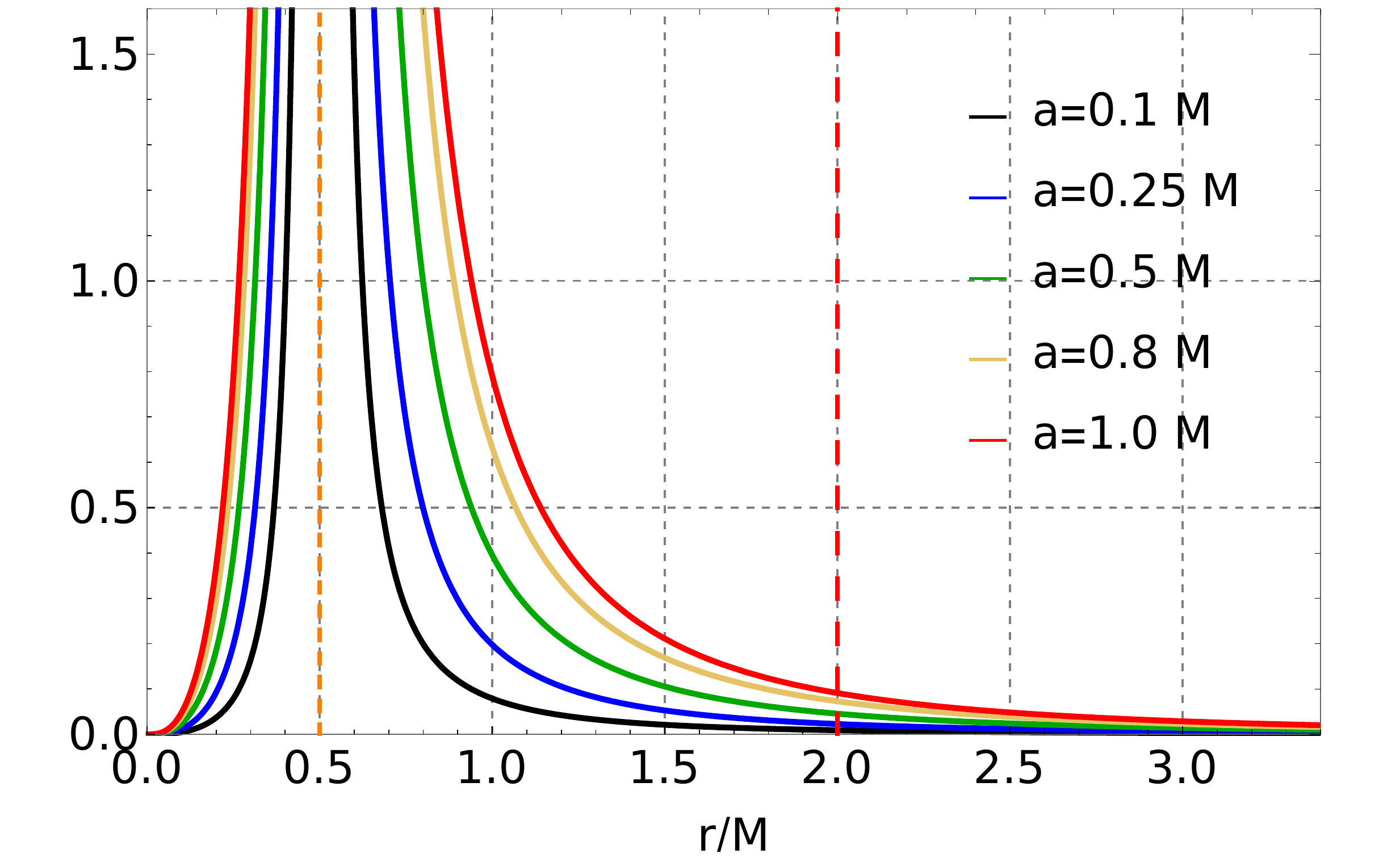}
    \end{subfigure}
    \\
    \centering
    \begin{subfigure}[t]{0.5\textwidth}
        \centering
        \caption{scalar (PNJ), $\theta=0$}
        \includegraphics[width=2.6in]{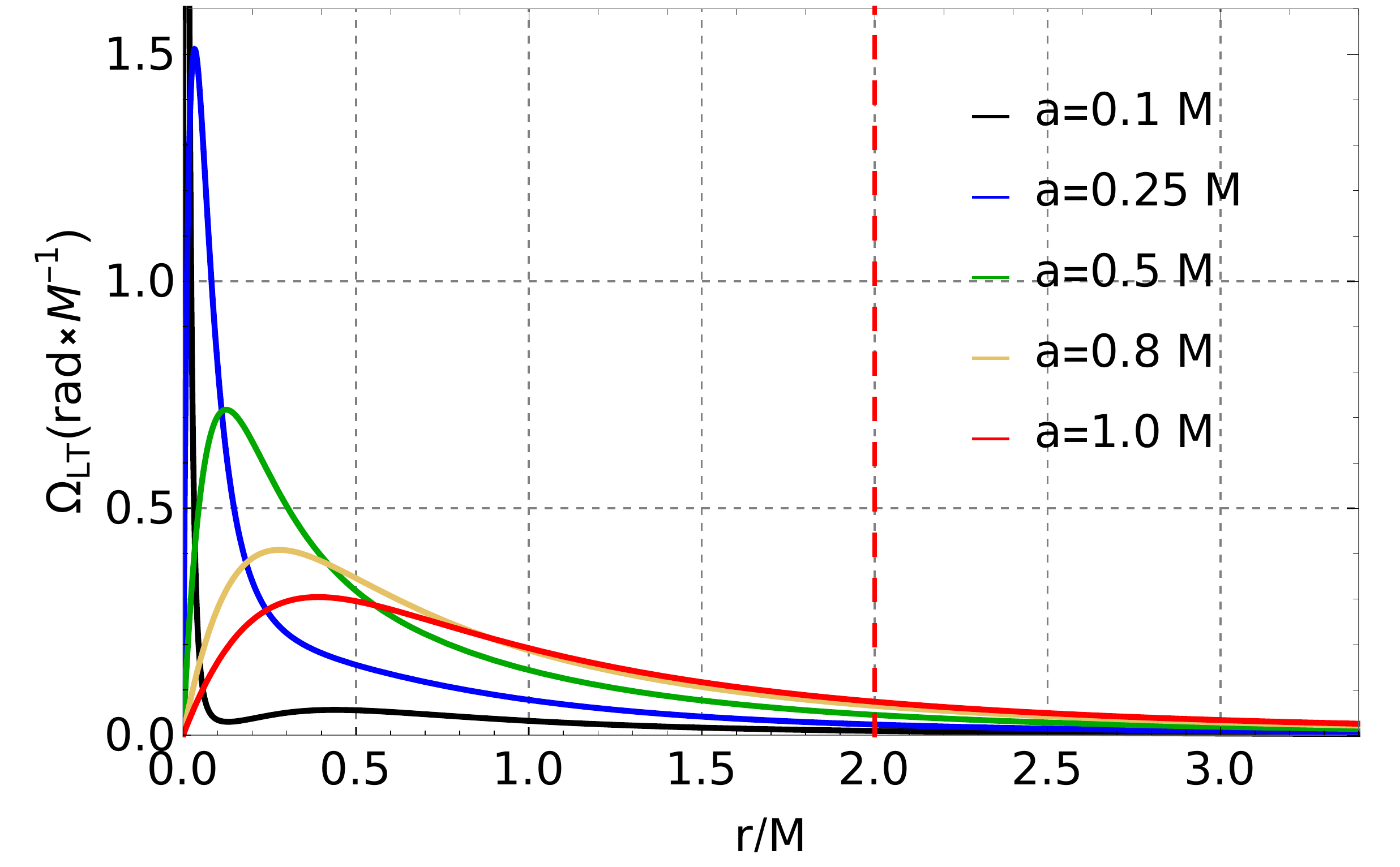}
    \end{subfigure}%
    ~ 
    \begin{subfigure}[t]{0.5\textwidth}
        \centering
        \caption{scalar (PNJ), $\theta=\pi/2$}
        \includegraphics[width=2.6in]{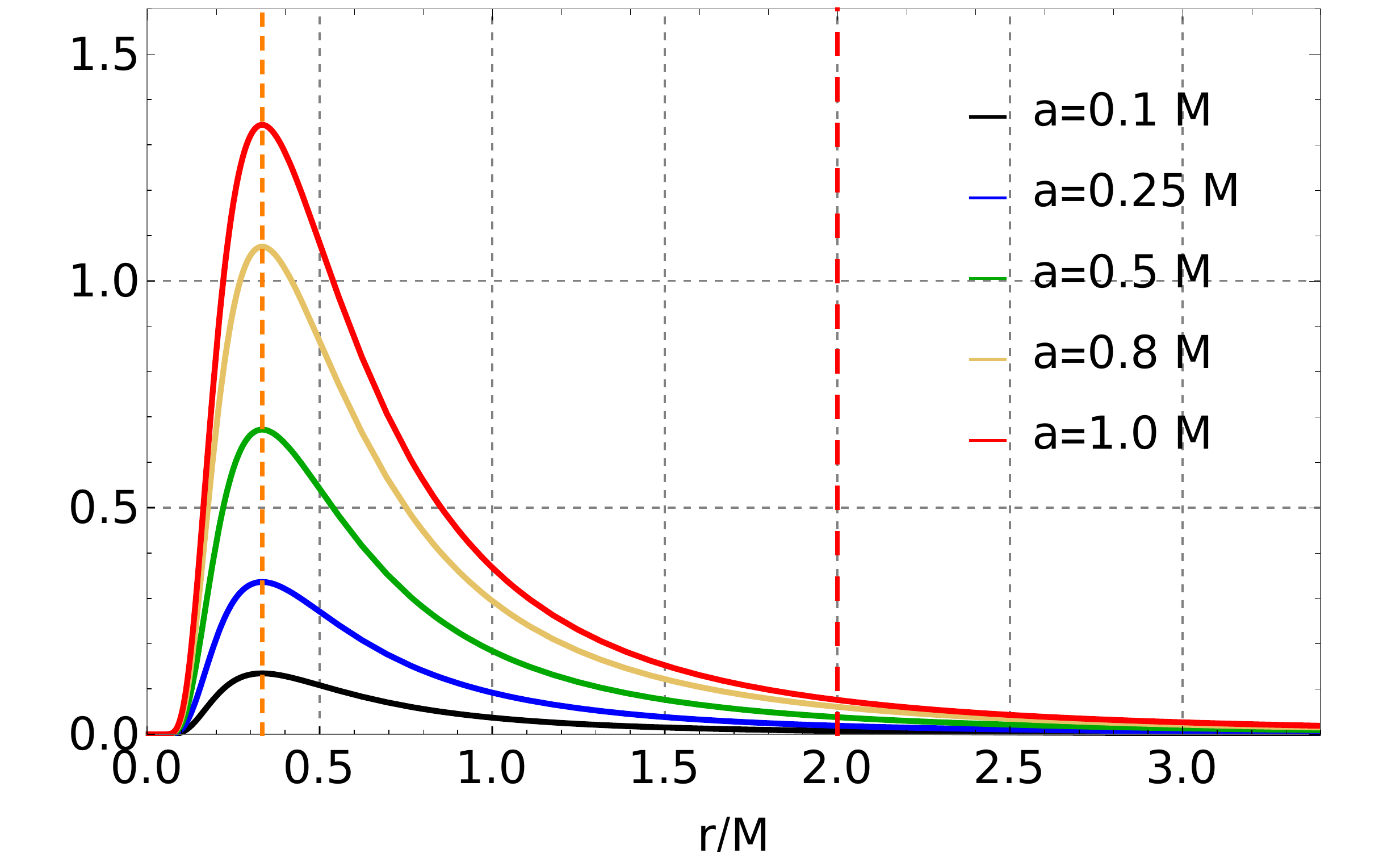}
    \end{subfigure}
    \caption{The Lense-Thirring precession frequency (in radians per unit
        time, $M$ in this case) as a function of normalized isotropic
        radial coordinate $r/M$, for the Kerr (top) and PNJ (bottom)
        metrics, for two values of the angle $\theta=0$
        (left) and $\theta=\pi/2$ (right), and for five values of the
        parameter $a$. The precession behavior in (b) at $r/M=0.5$ is singular as
        opposed to the scalar counterpart where it is finite.}
\label{fig:omegas}
\end{figure*}

\begin{figure}
	\center
	\includegraphics[width=8cm]{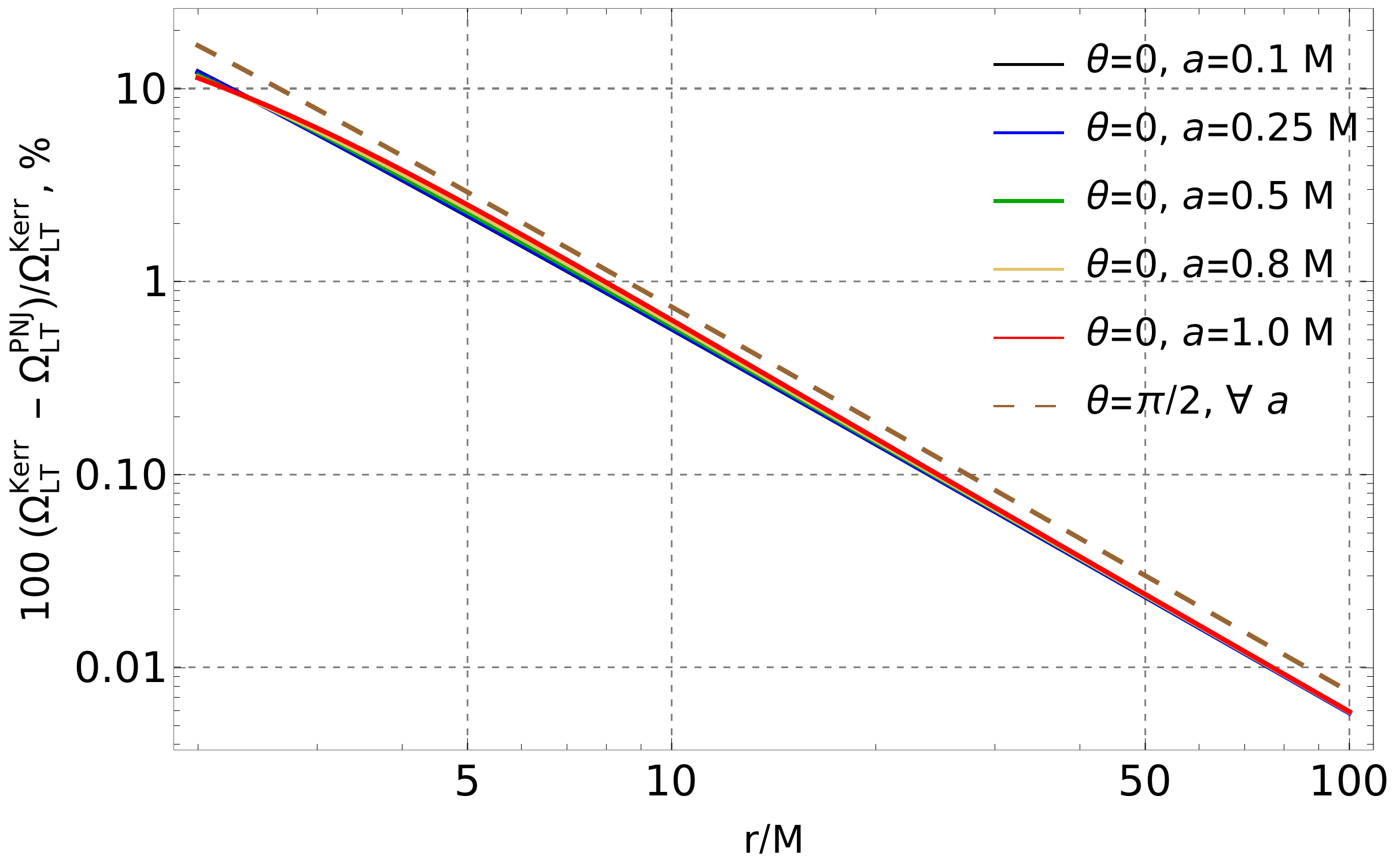}
    \caption{Relative difference in percents between Lense-Thirring frequencies
        $\Omega_{LT}^{Kerr}$ and $\Omega_{LT}^{PNJ}$ for larger values of
$r$ (log-log scale).}%
	\label{fig:omratio}%
\end{figure}

This final relation \eqref{LT_RAS} should be compared with its Kerr
counterpart rewritten in ``isotropic'' coordinates. For that, one
should substitute $R=r\left(1+\frac{M}{2r}\right)^2$ to get
$\Omega^{Kerr}_{LT}(R, \theta) \Rightarrow \Omega^{Kerr}_{LT}(r,
\theta)$. Some results of comparison of vacuum and scalar
Lense-Thirring effects within $r<r_g=2M$ are presented in
Fig.~\ref{fig:omegas}. For larger distances, the difference between
SNJ (Kerr) and PNJ cases decreases very rapidly, as shown in Fig.~\ref{fig:omratio}.

Due to axial symmetry planar orbits occur only for $\theta=0$ and
$\theta=\pi/2$. Manipulation in inclination angle $\theta$ shows that
the behavior of $\Omega^{Kerr}_{LT}$ in vacuum at $r=r_g/4=M/2$
becomes singular for $\theta=\pi/2$. 

In contrast, in scalar background at non-zero $r$ there are no
singularities in $\Omega_{LT}^{PNJ}$ which behaves monotonically and
increases with the growth of $a$ arriving, for the case (d), at its
maximum at $r\approx 0.32M=0.16r_g$ (orange vertical line) when moving
in the equatorial plane $\theta=\pi/2$. At scales $r>r_g=2M$ the
vacuum and scalar background precession effects practically coincide. 

\section{Kerr-Schild \`a la Yang-Mills}%
\label{sec:double_copy}

The simplest complex shift by an imaginary constant $z \rightarrow z +
ia$, applied to the Schwarzschild metric in Cartesian coordinates,
proves to be equivalent to the NJ transformation $R' = R
+ia\cos\theta$ (see Ref.~\refcite{arkani-hamed_kerr_2020}). Then, the Newtonian
potential function entering the Schwarzschild metric transforms
directly into the Kerr effective potential function, and this is a key
point of the NJ approach:
\begin{equation}
\varphi(R) = \frac{M}{R} \quad \Rightarrow  \quad \varphi(R, \theta,
a) = \frac{M}{2} \left(\frac{1}{R'} + \frac{1}{\bar{R'}} \right) =
\frac{M R}{R^2 +a^2 \cos^2\theta}.
\label{fifi}
\end{equation}
These expressions coincide with those following from the bimetric
Kerr-Schild form of vacuum solution \cite{kerr_republication_2009}:
\begin{equation}
    g_{\mu\nu} = \eta_{\mu\nu} + 2 \varphi k_\mu k_\nu, \qquad k^\alpha k_\alpha = 0,
    \label{KS}
\end{equation}
with functions $\varphi = \varphi(R)$ or $\varphi = \varphi(R, \theta,
a)$, correspondingly. Unlike in vacuum, in scalar background case the
physical scalar fields entering the metrics \eqref{Papa},
\eqref{rotPap} prove to be functionally the same as in \eqref{fifi}
but represent solutions of the Klein-Gordon equation \eqref{potRot}.

The Kerr-Schild representation of the vacuum solutions \eqref{KS}
linearizes the Ricci tensor, and as a result proves to be
closely related to the Yang-Mills approach in linear
U(1)-invariant regime through the so-called Kerr-Schild double
\cite{arkani-hamed_kerr_2020} and single \cite{alawadhi2021single}
copy maps of Yang-Mills fields.
 
The indicated similarity between the gauge and gravity theories
follows from correspondence between the gauge field $ F_{\mu\nu}$ and
the curvature  tensor $ R_{\mu\nu\lambda\delta}$. By analogy with
self-dual Yang-Mills presentation (which admits a wave physical
interpretation),
\begin{equation}
    F_{\mu\nu} = \frac{i}{2} \epsilon_{\mu\nu\rho\sigma}F^{\rho\sigma},
\end{equation}
the equations for self-dual gravity imply
\begin{equation}
    R_{\mu\nu\lambda\delta} = \frac{i}{2}
    \epsilon_{\mu\nu\rho\sigma}{R^{\rho\sigma}}_{\lambda\delta}.
\end{equation}
These might be analyzed in terms of the Kerr-Schild metric (cf.
\eqref{KS}), $g_{\mu\nu} = \eta_{\mu\nu} + 2 \kappa h_{\mu\nu} =
\eta_{\mu\nu} + 2 \kappa \varphi k_\mu k_\nu$, if one includes into
consideration a symbolic wave, the `graviton' $h_{\mu\nu}$. The null
character of $k_\mu$ is crucial for the linearity of the metric with
respect to scalar field $\varphi$, which greatly simplifies things.
Indeed, the Kerr-Schild form of the Kerr metric is chosen in analogy
with the linearized presentation of the propagation of perturbed metric.
However, the Kerr-Schild metric is not linearized but exact. The null
character of $k_\mu$ allows to single out the scalar function $\phi$ which
proves to be of the same functional form as the scalar field. On the
other hand, exactly this function $\phi$ is related
to the Yang-Mills zero component of the linearized Yang-Mills vector
potential \cite{gurses2018classical}.

Since the vacuum Einstein equations $R_{\mu\nu} = 0$ correspond, for
stationary regime, to the Yang-Mills equations
\begin{equation}
    \partial_\mu F^{\mu\nu} = \partial_\mu \left( \partial^\mu A^\nu -
    \partial^\nu A^\mu \right) = 0,
\end{equation}
the link between the gauge and Riemannian approaches might be
represented via the following ansatz (see, e.g., Refs.
\refcite{arkani-hamed_kerr_2020,gurses2018classical,Luna}):
\begin{equation}
    A^\mu = \varphi k^\mu.
\end{equation}
Then the comoving scalar potential function $\varphi$ (entering the
given ansatz and calculated first in Cartesian Kerr-Schild
coordinates \cite{kerr_gravitational_1963}) proves to be again of
familiar form (in appropriate units) \cite{Luna}: $ \varphi(R,\theta,
a) = MR/{\rho^2}$, as we already indicated before, see \eqref{fifi}
and \eqref{potRot}. 

So, the appearance of the given potential functions not only is
related to but, rather, might be explained by the correspondence with
non-gravitational Yang-Mills forces of nature. 

Thus, the connections between the Kerr-Schild representation of the
Kerr solution and stationary linear Yang-Mills fields were possible
due to linear character both of the Kerr-Schild algorithm and of the
corresponding Ricci-flat EE. The latter, in its turn, has origin in
the linearity of the field equations in curvature coordinates
(as commented in Section \ref{sec:genuine_reason}) with subsequent
application of the Ricci-flatness-preserving NJ algorithm. 

Importantly, in the case of scalar background, the Newman-Janis
algorithm applied to isotropic exponential metric leads to physical
scalar potential \eqref{potRot} of the same functional form as in the
Kerr-Shield vacuum case which might indicate common nature.

\section{Exact approach}
\label{sec:exact}

When one applies the exact rotational coordinate transformation
(hereafter, RCT)
\begin{equation}
\phi \to \phi - \Omega(r)t,
\label{omg}
\end{equation}
with a relevant angular velocity $\Omega(r)$
and corresponding exact differential
\begin{equation}
d\phi \to d\phi - \Omega(r)dt - t {\Omega}'(r)dr, 
\label{domg}
\end{equation}
to the Schwarzschild interval, 
one obtains an exact differentially rotating solution
(hereafter, RCTS, with $S$ for Schwarzschild) as an alternative to the special
algebraic Kerr (SNJ in our terms)
solution, which might be used for evaluation of 
rotational effects. So, we get the following non-stationary representation of the Schwarzschild solution:
\begin{equation}
	ds^2_S = N^2(r) dt^2 - \psi^4(r)
	\left( dr^2 + r^2 d\theta^2 +
	r^2 \sin^2\theta (d\phi - \Omega(r)dt - t
    {\Omega}'(r)dr)^2\right),
\label{SchwRot}    
\end{equation}
with coefficients $N(r)=Z(r)/\psi(r)$, $Z=1-\frac{M}{2r}$, $\psi = 1+\frac{M}{2r}$ defined in accord with the
standard Schwarzschild metric written in isotropic coordinates, or, in matrix form, 
\begin{equation}
{g}_{\mu \nu}^S = \psi^4  r^2\sin^2\theta  \left(
\begin{array}{cccc}
 B_S & - t    \Omega \Omega' & 0 &     \Omega \\
 - t    \Omega \Omega' & D_S & 0 &  t    \Omega' \\
 0 & 0 & -\csc^2\theta & 0 \\
     \Omega &  t    \Omega' & 0 & -1  \\
\end{array}
\right),
\label{SCovRot}
\end{equation}
where
$$ B_S = \frac{Z^2}{\psi^6  r^{2}\sin^{2}\theta}  - \Omega^2, 
\quad
   D_S = - \left( \frac{1}{r^2  \sin^2\theta} + t^2  \Omega'^2\right). 
$$

Note that the solution \eqref{SCovRot}, in
general, might be satisfied with different $\Omega(r)$ (including the case \,$\Omega=const$).
We adopt the asymptotic condition $\Omega(r) \to 0$ when
$r\to\infty$. 

Then from \eqref{SchwRot}-\eqref{SCovRot} substituted into
standard relations \eqref{GenPrec0} and \eqref{magnitudeLT} one
obtains, without approximations, the Lense-Thirring-type precession formula for
differentially rotating frame in vacuum: 
\begin{equation}
    \Omega_{LT}^S =  A(r,\theta) \sqrt{ P^2(r, \theta) + (\Delta^S_{LT})^2} ,
\label{LTSchwarz}
\end{equation}

\noindent with singular points  corresponding to the zeros of the denominator of
\begin{equation}
 A(r,\theta) = \frac{a M r \psi}{r^4 Z^2 \psi^6 -4 a^2 M^2 \sin^2\theta  }.   
 \label{art}
\end{equation}
Expanding the notation further,
\begin{equation*}
    P^2(r,\theta) = \sin ^2\theta \left(\frac{12 a^2 M^2 \sin^2\theta }{r^4 \psi^7}+\psi\right)^2+4 Z^2 \cos^2\theta, 
\end{equation*}
and the only non-stationary term, which is negligible if $t \Omega'\ll1$,
\begin{equation}
    \Delta^S_{LT} = t \Omega' Q,
\end{equation}
\begin{equation*}
    Q(r,\theta) =  Z r \sin{2\theta} ,   
\end{equation*}
where, according to Ref.~\refcite{cohen67}, we accept the differential rotation law:
\begin{equation*}
\Omega(r) = \frac{2 a M}{r^3 \psi^{6}},  
\quad
\Omega'(r) = -\frac{6aMZ}{r^4\psi^7}=  -\frac{6 a M
\left(1-\frac{M}{2r}\right)}{r^4\left(1+\frac{M}{2 r}\right)^7}.
\label{OmgSchw}    
\end{equation*}
Note that \eqref{LTSchwarz} is stationary for $\theta=0$ and
$\theta=\pi/2$, since $Q = 0$ in these cases. In accord with \eqref{art}, assuming $M=1$, $\theta=\pi/2$, the value
of $\Omega^S_{LT}$ has two singular
points floating (with $a$ varying from  
$a = 0$ to $a = 1$) in the interval from $r_1=r_2=r_g/4$ to $r_1=0.229$ and $r_2=1.091$, correspondingly.
Also, $\Omega^S_{LT}$ changes sign between
the singular points.

\begin{figure}
	\center
	\includegraphics[width=8cm]{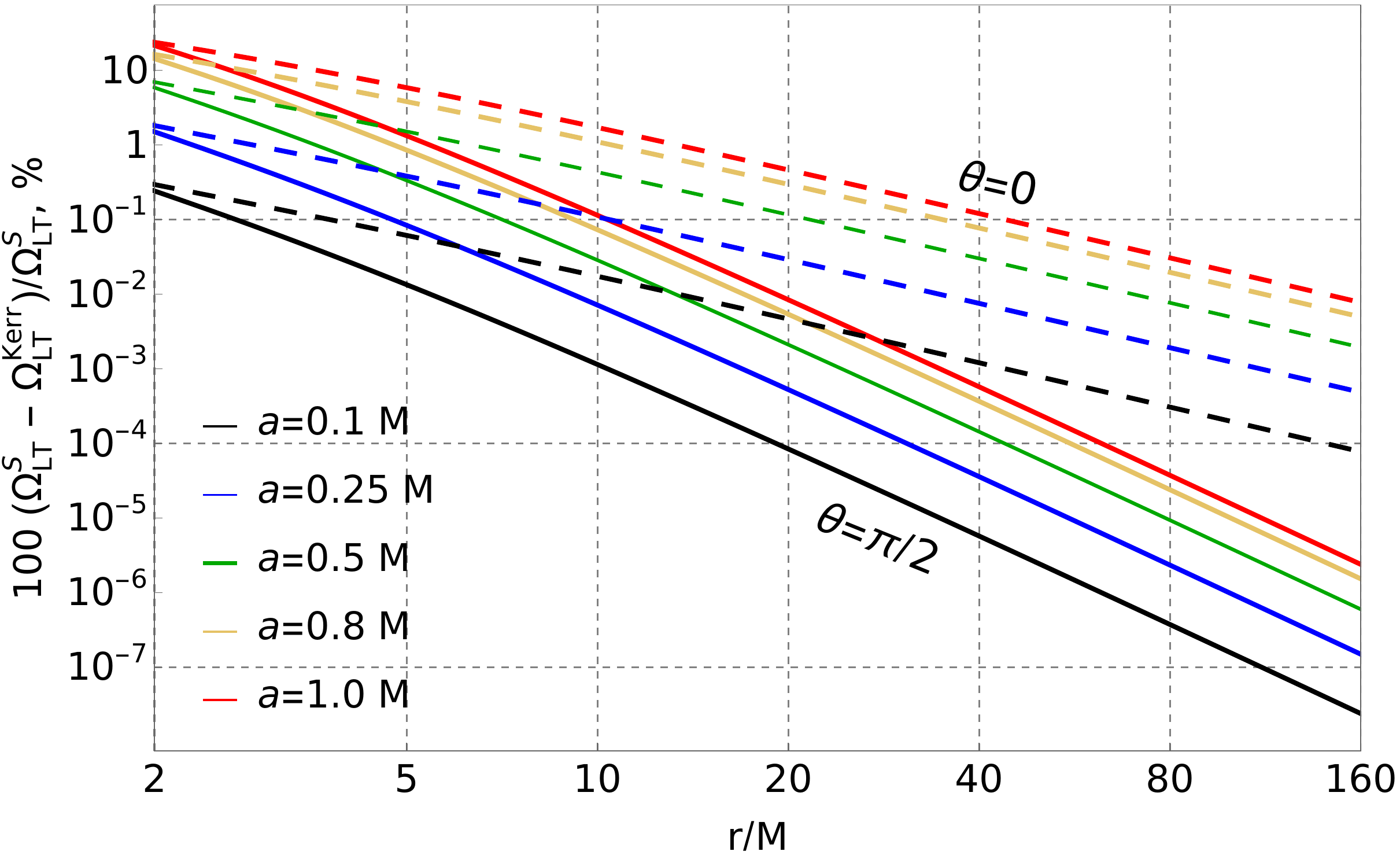}
    \caption{Relative difference in percents between Lense-Thirring frequencies
        $\Omega_{LT}^{S}$ and $\Omega_{LT}^{Kerr}$ for larger values of
$r$ and for two values of polar angle $\theta=0$ (dashed) and
$\theta=\pi/2$ (solid), in log-log scale.}%
	\label{fig:omratio_s_kerr}%
\end{figure}

It is remarkable that the Lense-Thirring effects both in the Kerr case
\cite{mm18} and for RCTS in stationary regime for $\theta=\pi/2$,
though very distinct analytically, might practically coincide (see
Fig.~\ref{fig:omratio_s_kerr}). So, as a typical example, the
numerical juxtaposition of \eqref{LTSchwarz} with the corresponding
expression for the Kerr metric (reduced to isotropic-like
coordinates\cite{mm18}) shows that the difference is essential only at
very small scales as a consequence of the initial coordinate
singularity at $Z=0$. 

For example, when $\theta=\pi/2$ and $a=0.25$ (blue solid line)
already at $r = r_g$ the distinction between $\Omega_{LT}^S$ and
$\Omega_{LT}^{Kerr}$ is $1.5\%$, for $r=3r_g$ it is
$0.045\%$, and for $r=10r_g$ it is  $0.00053\%$, etc., with
$\Omega_{LT}^S$ (RCTS) values slightly prevailing over
$\Omega_{LT}^{Kerr}$. Such almost negligible difference
between the Kerr and RCTS effects is quite remarkable. Thus, since we
are dealing with the coordinate transformation from the Schwarzschild
metric, the Lense-Thirring phenomenon should be treated, similarly to
that of Coriolis, as simply a coordinate effect.

For values of $\theta$ between $0$ and $\pi/2$,
$\Omega_{LT}^S$ is non-stationary due to time-dependence
of the metric, and should be studied separately.

\section{Conclusion and outlook}

We have analyzed the problem of application of the Newman-Janis
algorithm to scalar background in general relativity in antiscalar
regime, and in comparison with vacuum case. 

The resulting rotational PNJ metric \eqref{rotPap} (produced with that
algorithm from the exponential Papapetrou solution \eqref{Papa})
contains new rotational potential. Being the solution of the
Klein-Gordon equation this potential should be considered as physical.
It proves to be exactly of the same form as scalar function arising in
the bimetric Kerr-Schild approach in vacuum which, in its turn,
coincides with the Yang-Mills potential. 

This point might suggest that the Kerr spacetime (generated by the NJ
algorithm) should be interpreted as a sort of pregeometry for certain
(linearized) Yang-Mills fields. In such a case the PNJ metric appears
as horizonless (due to presence of scalar field) counterpart of that
spacetime.

The splitting between the geometric and material parts of the EE,
appreciable at small radial scale, is not so unexpected because this
metric makes the resulting scalar EMT incompatible with the
Hawking-Ellis energy condition \eqref{HE}. In such case the relevant
Einstein tensor exhibits anomalous structure and breaks the 
 compatibility with scalar EMT observed in static regime. 

Nevertheless, beyond the gravitational radius scale the effect of such
splitting becomes negligible and practically has no impact onto
distant observations. At the same time, the metric \eqref{rotPap} by
itself proves to be self-consistent and self-sufficient in the scalar
sector. 

The latter point deserves clarification. Indeed, as known, the
variational problem based on the Ricci scalar action is not well-posed
due to the necessity to fix, apart from the metric, its normal
derivatives on the boundary which, in general, might be incompatible
with the EE proper. This circumstance is discussed in detail in
Ref.~\refcite{paddy22}, where a radical workaround is proposed based
on the application of the path integral method to the action, with
subsequent exclusion of the metric from the category of dynamical
variables. 

A similar situation with the diminished significance of the metric
also arises within GR with scalar background. Then the scalar field
$\phi(x^\mu)$ becomes in fact the only dynamical variable when the
metric depends on coordinates exclusively through that field:
$$g_{\mu\nu}(x^\mu) \, \rightarrow \, g_{\mu\nu}(\varphi(x^\mu)).$$
So, the metric proves to be induced by scalar field and generates the
final self-sufficient form of EMT producing all testable observational
effects. Thus, in stationary case we get a metric emerging with the NJ
algorithm inside the scalar sector, but, unlike in static case, not
being an exact solution of the EE proper. 

Another example of such situation indicated in Ref.~\refcite{paddy22}
is related to the configuration with $\varphi=const$ satisfied by
${T^{\mu\nu}}_{;\nu}=0$ for massive scalar field which is not,
however, a solution of the EE.

The same situation is also found in the Kaniel-Itin model
\cite{hehl1998}, which contains the exponential metric, but scalar
field dynamics following from the 4-divergence of the modified EMT
does not comply with the EE as well.

We see that within the NJ framework one cannot also obtain the
stationary rotational metric compatible both with EE proper and with
the Hawking-Ellis criterion for the corresponding scalar EMT. 

Then, the role of the EE proper in this case is to supply the seed
exponential Papapetrou solution for subsequent generation of the
horizonless scalar counterpart of the Kerr metric being only
asymptotic solution of the EE and admitting also non-gravitational
interpretation of its origin. 

Purely gravitational variant, compatible both with the HE criterion
and the EE, is developed in Section \ref{sec:exact}, based on the
local rotational coordinate transformation which does not suggest
introduction of any new potentials. The price for the exact solution
thus obtained is, in general, its non-stationarity. At the same time,
for stationary regime $\theta=\pi/2$, direct calculation and
juxtaposition with the case of the Kerr metric yields practically
coinciding results for the Lense-Thirring effect which allows to
interpret it, similarly to that of Coriolis, as a coordinate effect,
as opposed to linearized dragging and gravitomagnetic interpretations.

On the whole, apart from the HE condition, the applicability of the EE
crucially depends on how exactly the rotation is introduced into
spacetimes. In fact, there are only very restricted ways to do that.
Here we had applied the Newman-Janis approach to scalar background,
and also exact rotational transformation with respect to the vacuum
Schwarzschild case as a viable alternative to Kerr. In a subsequent
work, we apply another known universal Brill-Cohen method
\cite{Brill1966} based on the differential rotation ansatz, as well as
relevant prolongation of the exact transformation approach to rotation
in scalar background.

\appendix

\section{$tt$-components of the EMT and Einstein
tensor for the PNJ metric}
\label{appendix:gtt}

In antiscalar regime, the Einstein equations \eqref{EinEq} are $
G_{\mu\nu}=-8\pi T_{\mu\nu}^{sc}$. Here, for the metric  \eqref{rotPap}
specified by \eqref{potRot}, all calculation for the left and right
sides of the EE might be performed in exact form. 

For example, for the covariant component of the EMT
of minimal (anti)scalar field \eqref{EMT} one obtains: 
\begin{equation}
8\pi T_{tt}^{sc} = \frac{M^2 \left(a^4+a^2 \left(a^2-3 r^2\right) \cos{2
		\theta}-a^2 r^2+2 r^4\right)}{2 \rho^8 Z^2}.
\label{appTtt}
\end{equation}
At the same time the corresponding covariant $tt$-component of the Einstein tensor proves to be analytically different and much more cumbersome:
\begin{multline}
    G_{tt} = \frac{1}{32(a^2 + r^2) (\rho^2 Z)^4 } \times \\ 
    \Bigg\{  3 a^6  p_1\cos{6 \theta}+3 a^4  p_2\cos{4 \theta}+ a^2  p_3\cos{2 \theta}+ \frac{3}{2} p_4+\frac{3}{4} a^8 \cos{8 \theta} \\
- 2Z \Bigg[3 a^6  (p_1-2 M r)\cos{6 \theta}+3 a^4  \left(p_2-4 a^2 M
    r-8 M r^3\right)\cos{4 \theta}+ a^2  \left(p_3+6 a^4 M r\right)\cos{2 \theta}\\
    +\frac{3}{2}\left(p_4+8 a^6 M r+16 a^4 M r^3\right)+ 
\frac{3}{4} a^8 \cos{8 \theta}\Bigg]\\
+Z^2 \Bigg[3 a^6  (p_1-4 M r)\cos{6 \theta}+3 a^4  \left(p_2-4 a^2 M^2-8 a^2 M r+12 M^2 r^2-16 M r^3\right)\cos{4 \theta } \\
+a^2  \left(p_3-16 a^4 M^2+12 a^4 M r-16 a^2 M^2 r^2\right)\cos{2 \theta}\\
  +  \frac{3}{2} \left(p_4-\frac{1}{3} 8 a^6 M^2+16 a^6 M r+8 a^4 M^2 r^2+32 a^4
    M r^3+\frac{64}{3} a^2 M^2 r^4-\frac{64 M^2 r^6}{3}\right)\\
  +  \frac{3}{4} a^8 \cos{8 \theta}\Bigg] \Bigg\},
\label{appGtt}
\end{multline}
where we have used the following notations:
\begin{equation*}
    Z =e^{2\varphi(r,\theta, a)} =
    e^{\frac{2rM}{\rho^2}}, \quad \rho^2 =  r^2 + a^2 \cos^2\theta,
\end{equation*}
\begin{equation*}
    p_1 \left( r, a \right)  = 2 a^2 + 3 r^2,
\end{equation*}
\begin{equation*}
    p_2 \left( r, a \right)  = 7 a^4 + 18 a^2 r^2 + 12 ^4,
\end{equation*}
\begin{equation*}
    p_3 \left( r, a \right)  = 42 a^6 + 135 a^4 r^2 +144 a^2 r^4 + 48 r^6,
\end{equation*}
\begin{equation*}
    p_4 \left( r, a \right)  = \frac{35}{2} a^8 + 60 a^6 r^2 + 72 a^4 r^4 + 32
    a^2 r^6.
\end{equation*}
Note that apart from diagonal EMT components, there are non-zero covariant
$t\phi$- and $r\theta$-components rendering the EMT  non-diagonalizable with 
local Lorentz transformations. The same is true
for the contravariant form of the EMT which thereby does not satisfy the Hawking-Ellis
criterion \eqref{HE}.

\section*{Acknowledgments}
This research is funded by the Science Committee of the Ministry of 
Science and Higher Education of the Republic of Kazakhstan (Grants No. AP08052312 and No.
AP08856184). We thank Prof. Friedrich Hehl for drawing our attention
to the relevant paper on the Kaniel-Itin model.

\bibliographystyle{ws-ijmpd}
\bibliography{ref}

\end{document}